\documentclass[usenatbib,a4paper,times,fleqn]{mnras}
\usepackage{graphicx,natbib,times,amssymb,amsmath,pstricks}
\usepackage{aas_macros}
\usepackage{bm,url}

\newcommand{\eqnref}[1]{Eq.~\ref{#1}}

\defcitealias{bbb03}{BBB03}

\title[Does the PDF lead to the IMF?]{Does turbulence determine the initial mass function?}

\author[Liptai \& Price]{\parbox{\textwidth}{David Liptai$^{1}$\thanks{david.liptai@monash.edu}, Daniel J. Price$^{1}$, James Wurster$^{1,2}$ and Matthew R. Bate$^{2}$} \\
$^{1}$Monash Centre for Astrophysics (MoCA) and School of Physics \& Astronomy, Monash University, Clayton Vic 3800, Australia \\
$^{2}$School of Physics, University of Exeter, Stocker Rd, Exeter EX4 4QL, UK
}

\pagerange{\pageref{firstpage}--\pageref{lastpage}} \pubyear{2016}

\date{}
\begin{document}
\label{firstpage}
\bibliographystyle{mnras}
\maketitle

\begin{abstract}
 We test the hypothesis that the initial mass function (IMF) is determined by the density probability distribution function (PDF) produced by supersonic turbulence. We compare 14 simulations of star cluster formation in 50 solar mass molecular cloud cores where the initial turbulence contains either purely solenoidal or purely compressive modes, in each case resolving fragmentation to the opacity limit to determine the resultant IMF. We find statistically indistinguishable IMFs between the two sets of calculations, despite a factor of two difference in the star formation rate and in the standard deviation of $\log(\rho)$. This suggests that the density PDF, while determining the star formation rate, is not the primary driver of the IMF.
\end{abstract}


\section{Introduction}
 
 
 Two decades of theoretical studies have established that a log-normal density Probability Distribution Function (PDF) is the defining characteristic of supersonic turbulence (e.g. \citealt{vazquez-semadeni94,ogs99,nordlundpadoan99,klessen00,kritsuketal07}; see review by \citealt{elmegreenscalo04}). In particular, numerous studies \citep[e.g.][]{pnj97,lemasterstone08,pfb11,molinaetal12} have shown that the density variance is proportional to the Mach number, giving
\begin{equation}
 \sigma_{\ln \rho}^{2} = \ln\left(1 + b^{2} \mathcal{M}^{2}\right), \label{eq:sigma}
\end{equation}
where $\sigma_{\ln \rho}$ is the standard deviation in the logarithm of the density (i.e. the `width' of the PDF), $\mathcal{M}$ is the root-mean-square (RMS) Mach number and $b$ is a constant of order unity related to the mixture of solenoidal and compressive modes in the velocity field \citep[e.g.][]{fks08,federrathetal10}. 

\citet{padoannordlund02} proposed that the PDF determines the IMF for low mass stars (M $< 1$ M$_{\odot}$), based on the observation that the IMF is also log-normal at the low-mass end \citep[e.g.][]{chabrier03,chabrier05}. Relating the PDF to the IMF is powerful because it enables analytic theories of star formation \citep[e.g.][]{krumholzmckee05,hennebellechabrier08,hennebellechabrier09,hopkins12,guszejnovhopkins15} that predict the IMF from the few parameters in Equation~\ref{eq:sigma}. Relating the IMF to the statistics of turbulence explains the universal nature of the IMF in the Milky Way \citep[e.g.][]{bcm10}, since nearby molecular clouds show supersonic motions with seemingly universal scaling relations \citep{zuckermanevans74,larson81,heyerbrunt04}. 
 
 Measurements of log-normal column density PDFs from extinction mapping \citep{lal06,lla08,lla10} lend support to a direct relationship between the PDF and the IMF. In particular, \citet{kainulainenetal09} showed that star-forming clouds differ from non-star-forming clouds by the presence of a power-law tail in the column density PDF at high densities, suggesting that self-gravity merely converts the high-density end of the PDF into stars. The measured mass function of `cores' also seems to mimic the stellar initial mass function, but shifted to higher masses, implying a one-to-one relationship between `cores' and `stars' with an efficiency factor of $\sim 0.3$ (e.g. \citealt{man98,testisargent98,luhmanrieke99,johnstoneetal00,all07,nutterward-thompson07,enochetal08,rathborneetal09,chabrierhennebelle10}). However, numerous studies have also cautioned or argued against a direct core mass function (CMF)-IMF relationship \citep[e.g.][]{ballesteros-paredesetal06,goodwinetal08,scb08,scb09}.

 Alternatively, \citet{bonnelletal97} and \citet{batebonnell05} proposed that the IMF is determined by `competitive accretion' between low mass fragments for a limited gas supply, with accretion truncated by the preferential ejection of low mass stars and brown dwarfs from unstable multiple systems. This was demonstrated in the star cluster formation calculations of \citet*{bbb03} (hereafter \citetalias{bbb03}). These were the first attempts to simulate the IMF `directly' by resolving the gravitational collapse to the opacity limit for fragmentation (the density at which radiation is trapped by dust, $\rho \approx 10^{-13}$g/cm$^{3}$, implying an increase rather than decrease in the Jeans mass with density, and hence the formation of a single hydrostatic object; \citealt{lowlynden-bell76,rees76}). Sink particles were inserted in the calculations once the opacity limit was reached, enabling simulation of the subsequent accretion up to the final stellar masses. This approach has had remarkable success at reproducing the observed IMF, with the most recent calculation by \citet{bate12} modelling the formation of 183 stars and brown dwarfs from a 500~M$_{\odot}$ cloud, finding an IMF statistically indistinguishable from the local IMF compiled by \citet{chabrier05}.
 
  While these simulations employ turbulent clouds, it is not obvious how the resultant IMF relates to the details of the initial turbulence. A subsequent study by \citet{bate09b} found no change in the IMF when the slope of the power spectrum of the initial turbulence was varied. In their simple model to explain the IMF produced by simulations, \citet{batebonnell05} invoke the PDF only indirectly, via a log-normal distribution of mass accretion rates. Nevertheless, a connection may still exist.
 
 Here we investigate the PDF-IMF connection by simulating star formation in two initially identical sets of model clouds, set up with either purely solenoidal or purely compressive initial velocity fields. If the PDF determines the IMF, then we expect the IMFs to differ, since the PDFs should be very different. If the IMF is more due to \emph{nurture} than \emph{nature}, the effect may be more minor. The main caveat to our study is that we assume impulsive rather than continuous turbulent driving.
 
  \citet{girichidisetal11} performed a related study, along with other variations in the initial conditions, and found that the shape of the IMF was unaffected by the type of turbulent driving. However, they simulated more massive and denser clouds (M = 100 M$_{\odot}$ and $R=0.1$ pc) and did not resolve to the opacity limit (sinks were inserted at a scale of 40~au, compared to 5~au employed here and in \citetalias{bbb03}). We also perform a statistical study with multiple realisations of the initial velocity field in each case, compared to their single realisation. \citet{lwh15} recently compared the effect of solenoidal vs. compressive forcing in star formation calculations, but focussed on smaller cores (M = 3 M$_{\odot}$; R = 3000 au), examining the effect on disc and binary fractions rather than the IMF.
  
  While this paper was under review, an important and complementary study to ours was published by \citet{bertelli-mottaetal16}, examining the correlation between the IMF and the statistics of turbulence using two sets of simulations where the turbulence was first driven to a steady state in a periodic box before `switching on' gravity. These authors varied the Mach number as well as the density of the cloud, using a total mass of either 5750 M$_{\odot}$ or 516 M$_{\odot}$ in a 10~pc$^{3}$ or 3~pc$^{3}$ domain, respectively. Their `high density' simulations were resolved only to a density of $1.6 \times 10^{-14}$ g/cm$^{3}$, one order of magnitude less than the opacity limit, with sink particle radii of 100~au. They found no correlation between the Mach number and the characteristic mass of the resulting IMF, concluding that the IMF is mainly determined by small scale processes such as disc formation and fragmentation and not by turbulence driven at the scale of the cloud. However, studying the role of initial conditions in a clump with decaying turbulence remains important since this may be closer to the situation in dense cores prior to the onset of stellar feedback.

 
\section{Numerical method}
\label{sec:methods}
 We use the \textsc{phantom} smoothed particle hydrodynamics (SPH) code \citep*{pricefederrath10,lodatoprice10,price12}. This is the first application of \textsc{phantom} to star cluster formation.

\subsection{Initial conditions}
 Aside from the initial velocity fields, our setup is identical to that in \citetalias{bbb03}: We set up a series of turbulent, spherical clouds, with $50 \text{ M}_\odot$ of gas of uniform density with diameter 0.375~pc. The corresponding initial free-fall time is $t_{\mathrm{ff}}=1.90 \times 10^6$ years. The minimum Jeans mass at the opacity limit is $M_\mathrm{min}\approx 0.0011 \text{ M}_\odot$. We use 3.5 million SPH particles, consistent with \citetalias{bbb03}, who showed that about 75 particles are required per $M_\mathrm{min}$ (see also \citealt{bateburkert97}). Particles were distributed in a uniform random distribution. We adopt code units with a length unit of $0.1{\rm pc}$, mass unit of $1 {\rm M}_{\odot}$ and time units such that $G=1$.
 
\subsection{Equation of state} 
 We adopt a barotropic equation of state $P=K\rho^\gamma$. Following \citetalias{bbb03}, we prescribe $\gamma = 1$ (i.e. isothermal) for densities lower than the opacity limit for fragmentation ($\rho=10^{-13}\text{ g cm}^{-3}$), $\gamma=7/5$ for $10^{-13}\text{ g cm}^{-3} < \rho < 10^{-10}\text{ g cm}^{-3}$ and $\gamma = 1.1$ for $\rho > 10^{-10}\text{ g cm}^{-3}$. We define the constant $K$ to be such that the sound speed is $c_s=1.84 \times 10^4$ cm s$^{-1}$ during the isothermal phase (i.e. 10 K assuming a mean molecular weight $\mu=2.46$) and in the $\gamma=7/5$ regime such that the pressure remains continuous when $\gamma$ changes. As discussed by \citet{bate09a}, using a barotropic equation of state over-produces low mass stars and brown dwarfs compared to observations, since the cold gas surrounding the protostars fragments too readily (c.f.~Fig.~\ref{fig:imfs}). Several groups \citep{bate09,bate12,offneretal09,krumholzetal10,commerconetal10} showed that this can be solved by modelling radiation in the flux-limited diffusion approximation. However, simulations with radiation are expensive, precluding the kind of statistical study we perform here, the radiation algorithm is not yet implemented in {\sc phantom}, and a barotropic equation of state is sufficient to answer the question of whether the PDF influences the IMF. We also ignore magnetic fields which change the star formation rate and perhaps also the IMF \citep{ogs99,hmk01,vkb05,tilleypudritz07,pricebate08,pricebate09,myersetal14}.

\begin{figure*}
\begin{center}
\includegraphics[width=\textwidth]{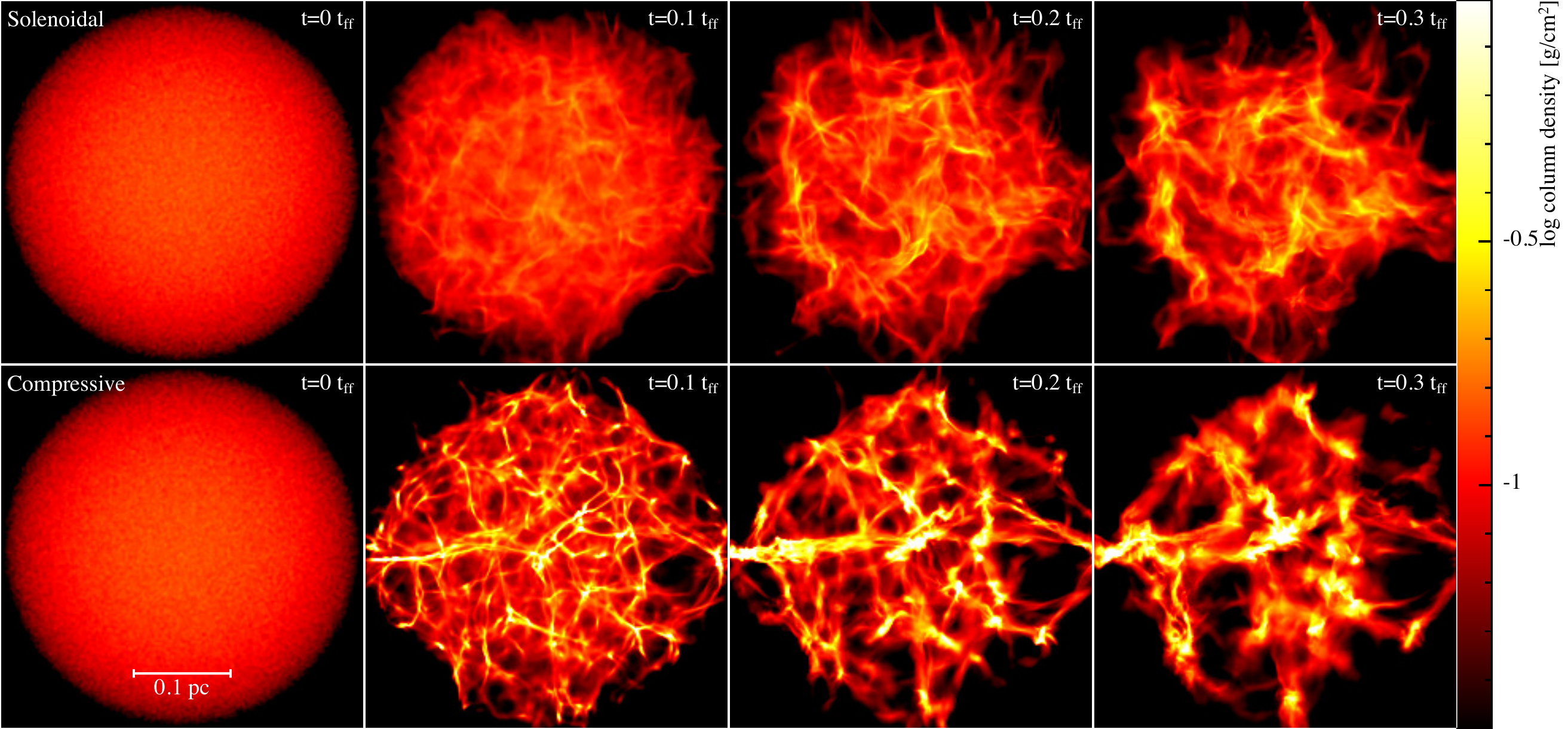}
\caption{Evolution of column density during the gravitational collapse of two example 50 M$_{\odot}$ molecular cloud cores with purely solenoidal (top) and purely compressive (bottom) initial turbulent velocity fields. The large scale structure of the clouds is very different, with the compressive case showing a factor of two increase in the standard deviation of $\log(\rho)$ compared to the solenoidal case as well as stronger shocks and a faster onset to star formation. To obtain enough statistics to determine the initial mass function, we perform simulations using 7 realisations of each type of driving, giving 14 simulations in total.} 
\label{fig:clouds}
\end{center}
\end{figure*}

\subsection{Velocity fields: solenoidal vs. compressive driving}
We impulsively drive turbulence in each cloud, as in \citetalias{bbb03}, by imposing an initial supersonic turbulent velocity field. The amplitude of the velocity fluctuations follow a power spectrum $P(k)\propto k^{-4}$ where $k$ is the wavenumber, in order to be consistent with Larson's scaling relation. We generate each field via a Fourier transform on a $64^3$ grid, which is then interpolated onto the SPH particles. The coefficient of each Fourier mode is drawn from a Rayleigh distribution with each mode also given a uniform random phase between $[-\pi,\pi]$. This is equivalent to sampling from a cylindrical bivariate Gaussian \citep*{dnp95}. 

To obtain a purely solenoidal velocity field, we take the curl of a vector field to produce a divergence-free velocity field.
Similarly for a purely compressive velocity field, we take the gradient of a scalar field to produce a curl-free field. We compute the gradients in Fourier space.
Velocities are normalised so that the initial kinetic energy is equal to the gravitational potential energy, giving an initial RMS Mach number of $\mathcal{M}=6.4$. We performed simulations using 7 realisations of the initial velocity field for each case (solenoidal or compressive), realised by changing the seed in the random number generator for the phases and amplitudes.

\subsection{Sink particles}
Following \citetalias{bbb03}, we introduce sink particles \citep{bbp95} when the central density of pressure-supported fragments reaches $\rho_s = 10^{-11}$ g cm$^{-3}$, two orders of magnitude higher than the opacity limit. Once $\rho_s$ is exceeded and sink formation conditions are satisfied, we replace gas particles within 5 au with a sink particle. Gas particles within 5~au are accreted if they pass checks for angular momentum and boundness, with their mass and momentum added to the sink. Gravity between sinks is softened within 4 au; gas particles are accreted without checks within this radius. 

\section{Results}
\label{sec:results}
\subsection{Column density evolution}
 Figure~\ref{fig:clouds} shows the evolution of column density from $t=0$ to $t=0.3 t_{\rm ff}$ (left to right) in two representative calculations, using solenoidal driving (top, as in \citetalias{bbb03}) and compressive driving (bottom). Shocks form quickly in both cases, due to the impulsive supersonic velocity field, but are stronger in the compressive case, driving the formation of large scale filaments after only $0.3 t_{\mathrm{ff}}$. For the solenoidal case, $\nabla \cdot \mathbf{v} = 0$ initially by definition, so there are no regions that initially promote collapse. 
 
\begin{figure}
\begin{center}
\includegraphics[width=\columnwidth]{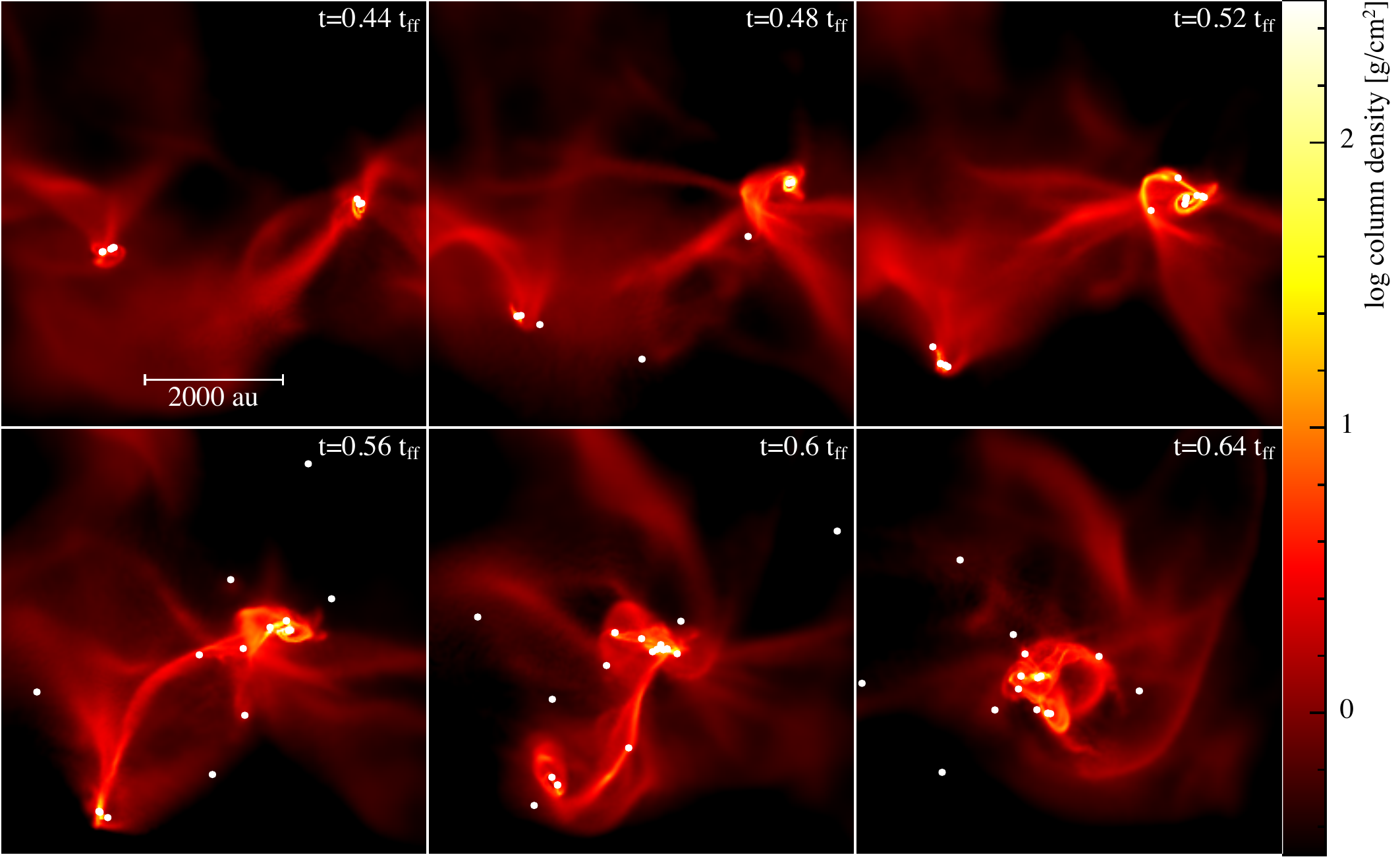}
\caption{Snapshots of the evolution after the onset of star formation, showing column density in a 0.03 pc $\times$ 0.03 pc inset for one of our compressively-driven clouds. The star formation process is similar in solenoidal clouds, but occurs later and at a slower rate.}
\label{fig:sf_core}
\end{center}
\end{figure}

 Figure~\ref{fig:sf_core} shows the subsequent small-scale fragmentation in the compressive cloud, with the first protostar formed after just $0.2 t_{\rm ff}$. The process in all other clouds appears visually very similar. Gas flows into dense cores along filaments \citep[e.g.][]{gomezvazquez-semadeni14,smithetal16,federrath16,kpk16}, feeding young protostars via accretion discs. The process is chaotic and dynamical, with close encounters between stars resulting in the destruction of accretion discs, and the ejection of smaller mass objects. Bound systems form and get destroyed by interactions on a very short time-scale. The stars live in a competitive environment, where those that grow in mass quickly stay in the dense regions and accrete further material, whilst ejecting lower mass objects.

\begin{figure}
\begin{center}
\includegraphics[width=0.7\columnwidth]{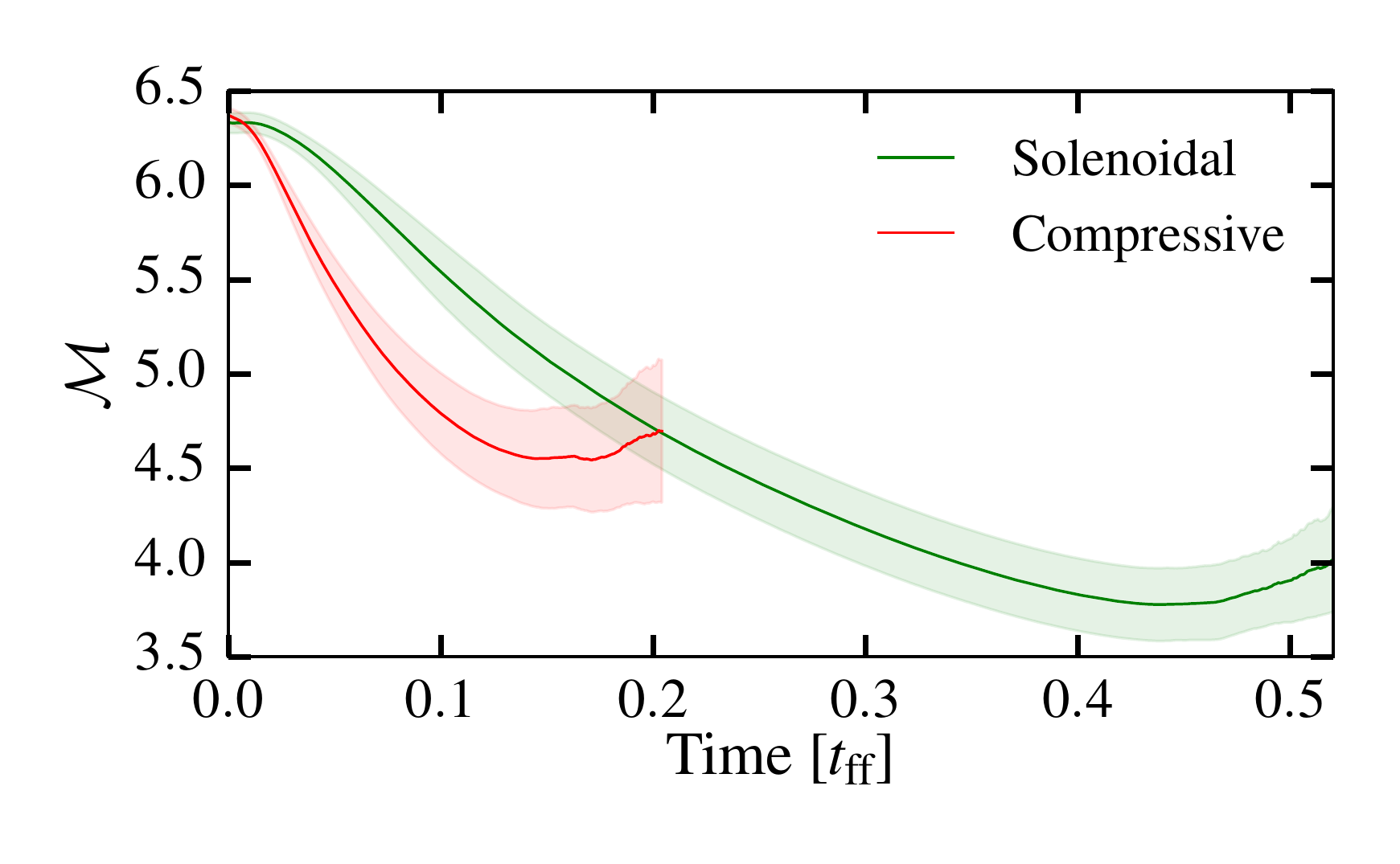}
\includegraphics[width=0.7\columnwidth]{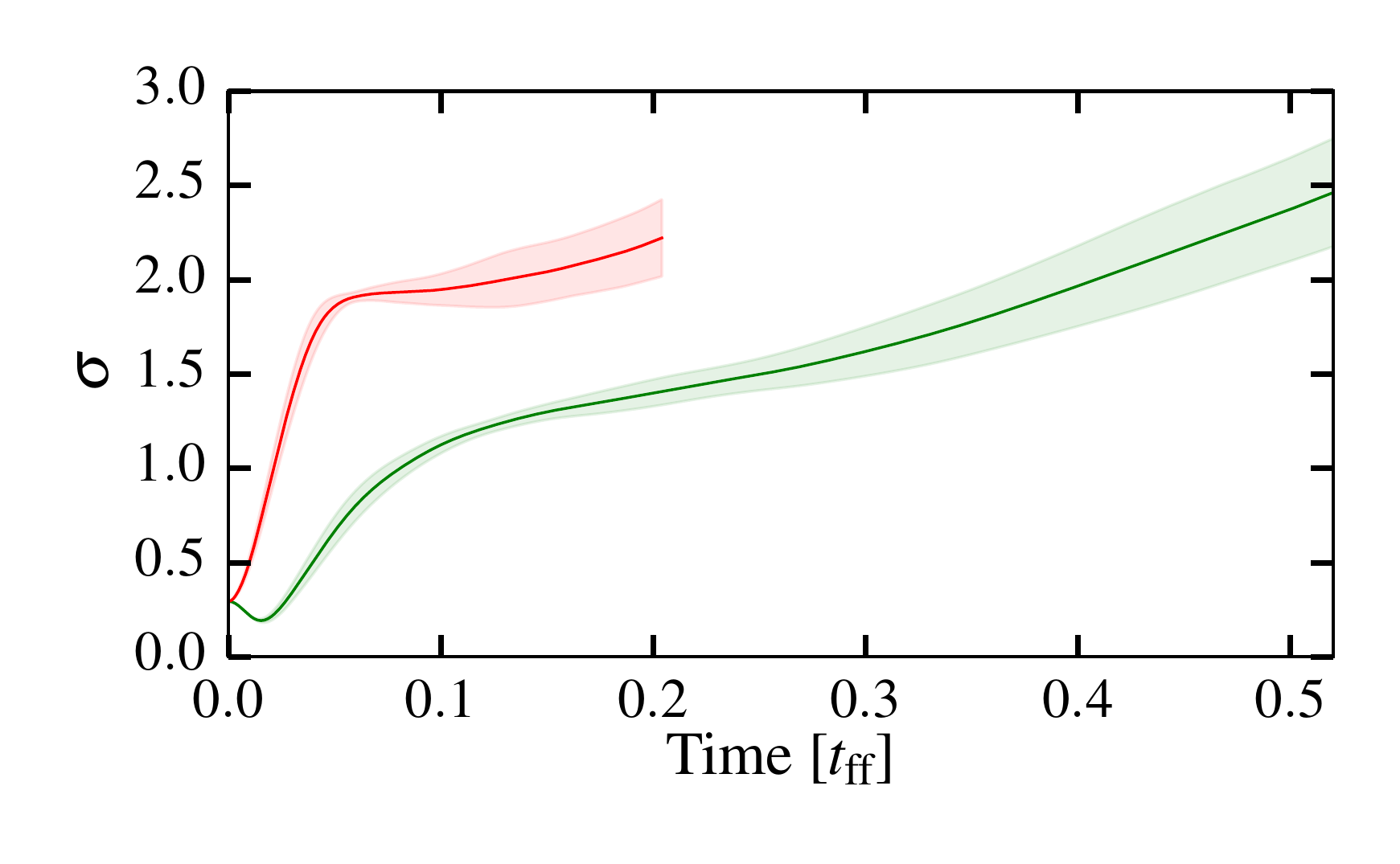}
\includegraphics[width=0.7\columnwidth]{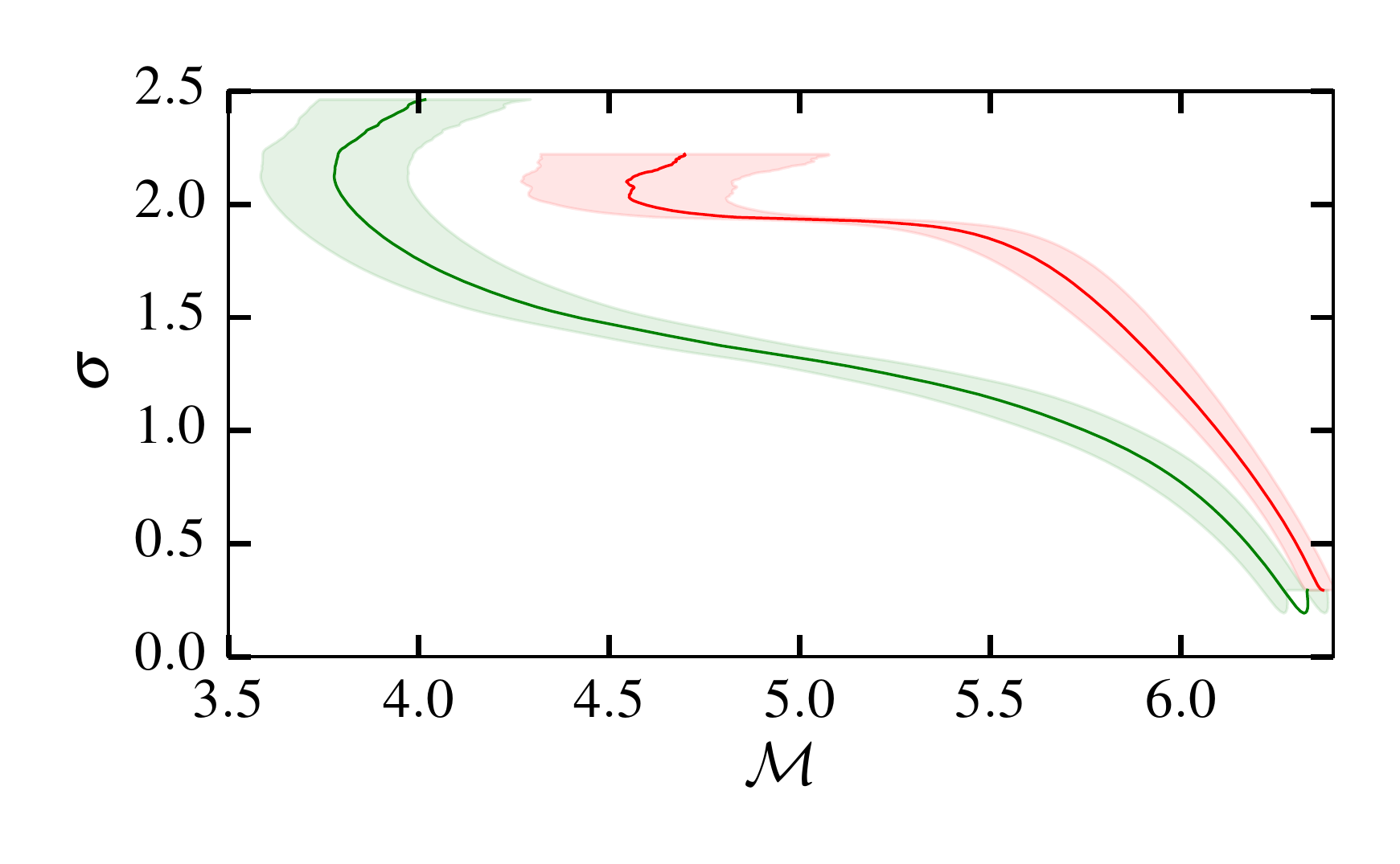}
\caption{The time evolution of the RMS Mach number $\mathcal{M}$ (top) and the mass-weighted standard deviation of the logarithm of density $\sigma_{\ln \rho}$ (middle). The lower panel shows the evolution in the $\sigma$-$\mathcal{M}$ plane. Solid lines show the mean over all 7 simulations of each type while the shaded error bars indicate the standard deviation between simulations.} 
\label{fig:pdf_stats}
\end{center}
\end{figure}

\subsection{Comparison of PDFs}
 We computed the density PDFs by binning the particles into 2000 bins equally spaced between $-10 < \log_{10}(\rho) < 10$ in code units. We then computed the standard deviation, $\sigma_{\ln \rho}$ by fitting a log-normal distribution to the PDF (using \verb+scipy.optimize.curve_fit+ in \textsc{Python}). Note that the PDF computed in this way is mass, rather than volume-weighted. Both volume and mass-weighted PDFs are expected to be log-normal when the equation of state is approximately isothermal \citep{pjn97,passotvazquez-semadeni98,scaloetal98,nordlundpadoan99,osg01}.
 
 Figure~\ref{fig:pdf_stats} shows the time evolution of the (mass weighted) RMS Mach number, $\mathcal{M}$, (top panel) and standard deviation, $\sigma_{\ln \rho}$, (centre panel) for our entire set of calculations, with the solid lines showing the mean from the 7 different simulations for each type of velocity field and the shaded region shows the 1$\sigma$ standard deviation. The bottom panel shows the evolution in the $\mathcal{M}$-$\sigma_{\ln \rho}$ plane. In both the solenoidal and compressive clouds $\mathcal{M}$ decays with time due to the dissipation of energy by shocks, reaching a minimum before rising again once bound structures have formed.
 
  Comparison of PDFs in decaying turbulence simulations is complicated by the time evolution of the velocity field. In our calculations the initial density field is uniform and the PDF thus develops in response to the initial turbulent velocity field. Since the clouds evolve on different time-scales, it is not particularly meaningful to compare their PDFs at the same time. Rather --- for the purposes of our study --- Equation~\ref{eq:sigma} suggests that they should be compared at the same RMS Mach number $\mathcal{M}$ so that the only difference is from the different mixing parameters $b$. 

  The lower panel of Figure~\ref{fig:pdf_stats} shows that the initial collapse of the cloud roughly corresponds to $\sigma_{\ln \rho} \lesssim 2$. Once $\sigma_{\ln \rho}$ reaches this value $\mathcal{M}$ rises again once fragmentation begins. Also, the PDF is no longer log-normal. We thus use the time interval where $\sigma_{\ln\rho} < 2$ to compare the density PDFs prior to the onset of star formation. The standard deviation of the PDFs is different not only at the same time early in the evolution of the cloud, but also at the same RMS Mach number.
  
\begin{figure}
\begin{center}
\includegraphics[width=\columnwidth]{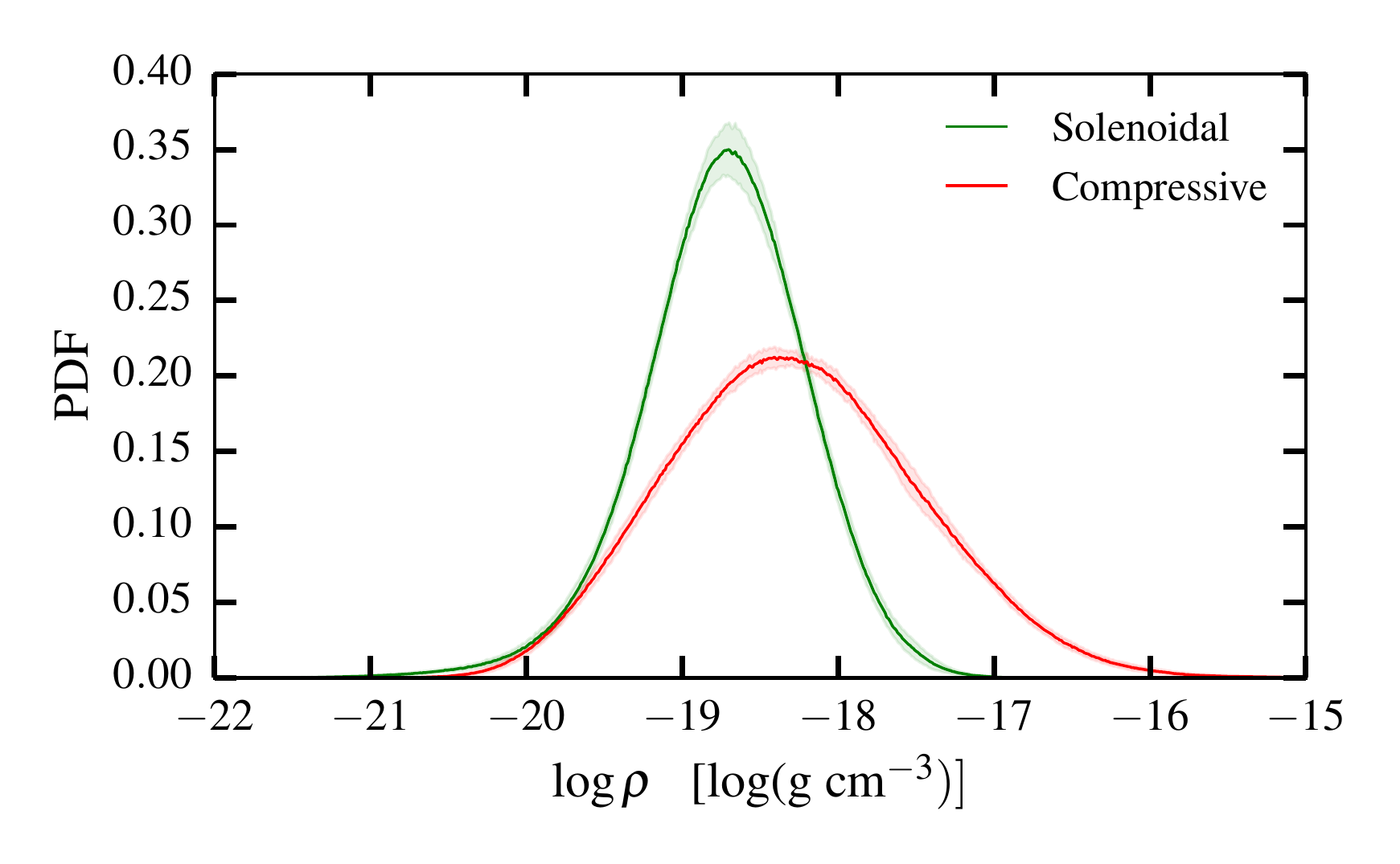}
\caption{Comparison of the mass-weighted density PDFs for the two types of turbulent driving, compared at the same RMS Mach number of $\mathcal{M}=5.5$. Solid lines show the mean over all 7 simulations of each type while shaded regions represent the 1$\sigma$ deviations between different realisations.} 
\label{fig:pdfs}
\end{center}
\end{figure}

Figure~\ref{fig:pdfs} shows the resultant PDFs computed at the time when all calculations have the same RMS Mach number of $\mathcal{M}=5.5$, which is when $\sigma$ differs most between the simulations. The difference in the PDF produced by compressive vs. solenoidal driving is similar to that shown by (e.g.) \citet{fks08} and \citet{federrathetal10}, except that we show the mass-weighted version. Compressive driving produces a broadening of the PDF caused by the collision of stronger shocks which in turn create larger variations in the density field. This demonstrates that our different choices of impulsive driving indeed drive significant differences in the density PDF prior to star formation.
 
\begin{figure}
\begin{center}
\includegraphics[width=0.8\columnwidth]{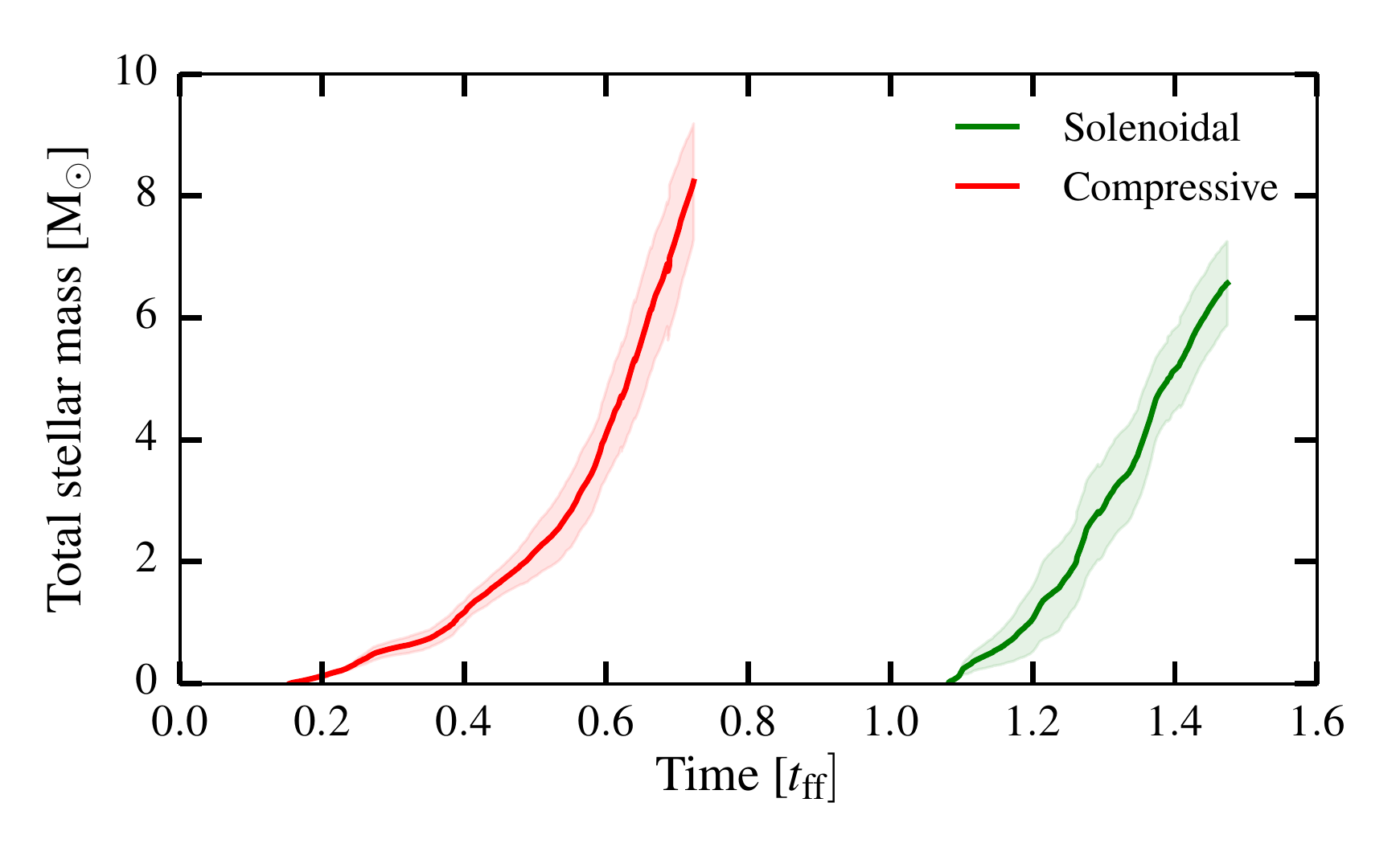}
\caption{Total mass in sink particles as a function of time for the two types of driving. The star formation rate is higher by a factor of two in the calculations employing compressive driving. The onset of star formation also occurs $\approx 0.9$ free-fall times earlier.} 
\label{fig:sfr}
\end{center}
\end{figure}

\begin{figure*}
\begin{center}
\includegraphics[width=\columnwidth]{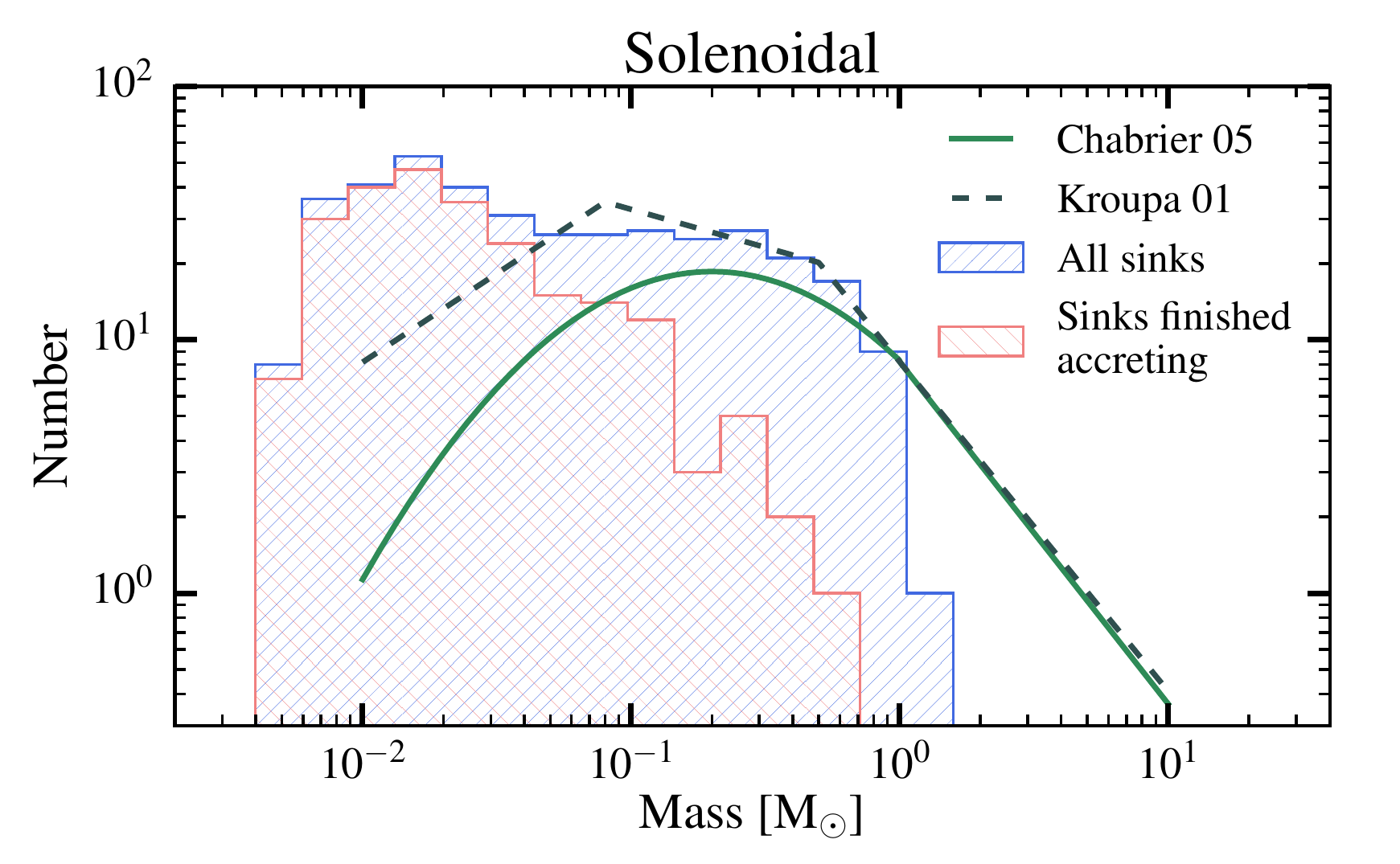}
\includegraphics[width=\columnwidth]{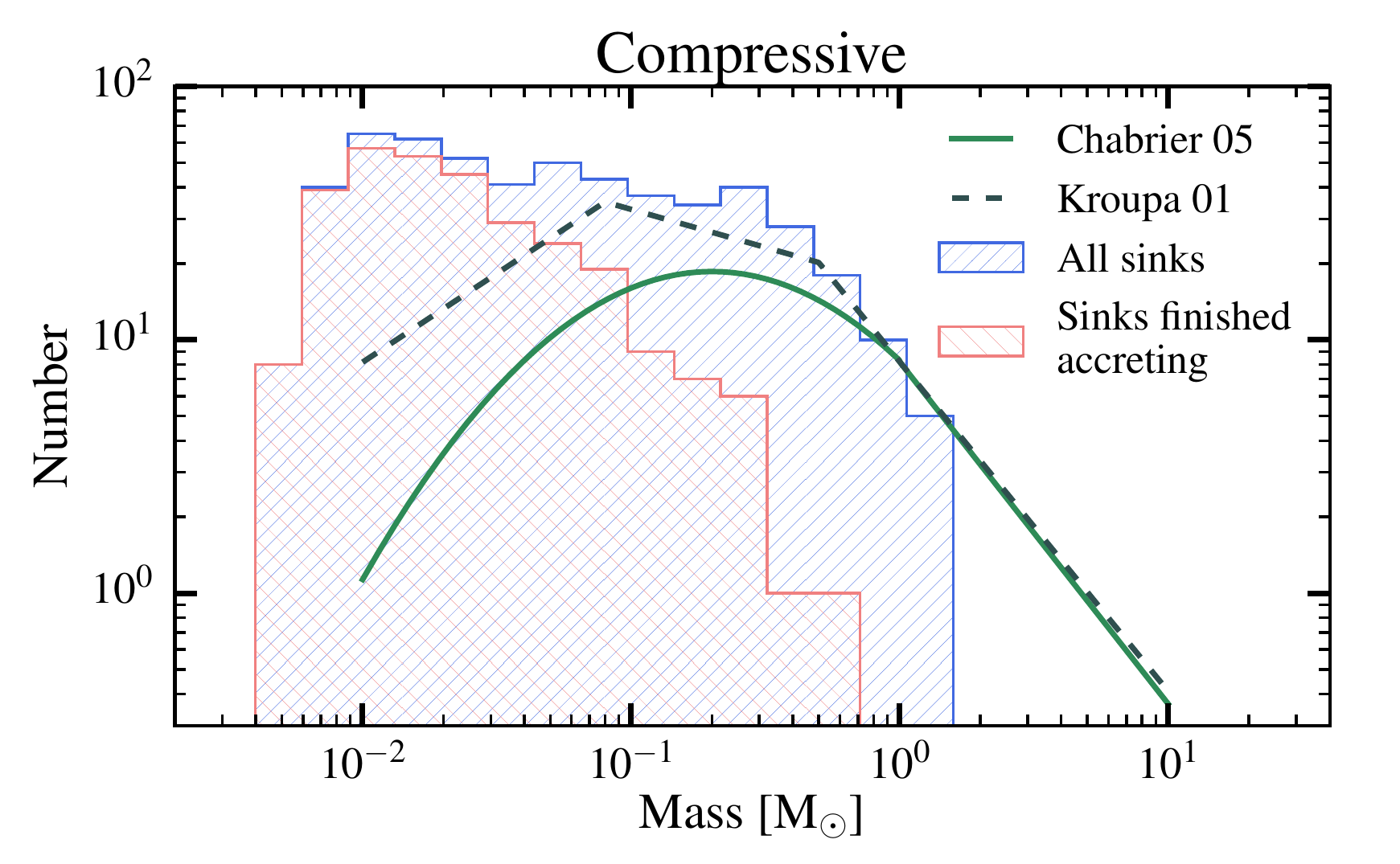}
\caption{Combined IMFs (blue and red histograms) from the 7 solenoidal (left) and 7 compressive (right) simulations. Solid/dashed lines show the empirically derived IMFs of \citet{kroupa01} and \citet{chabrier05} for comparison. While our simulations overproduce low mass objects, consistent with \citet{bate09a}, the IMFs with either solenoidal or compressive driving are statistically indistinguishable, suggesting no direct link between the PDF (Figure~\ref{fig:pdfs}) and the IMF.}
\label{fig:imfs}
\end{center}
\end{figure*}

\begin{figure}
\begin{center}
\includegraphics[width=0.9\columnwidth]{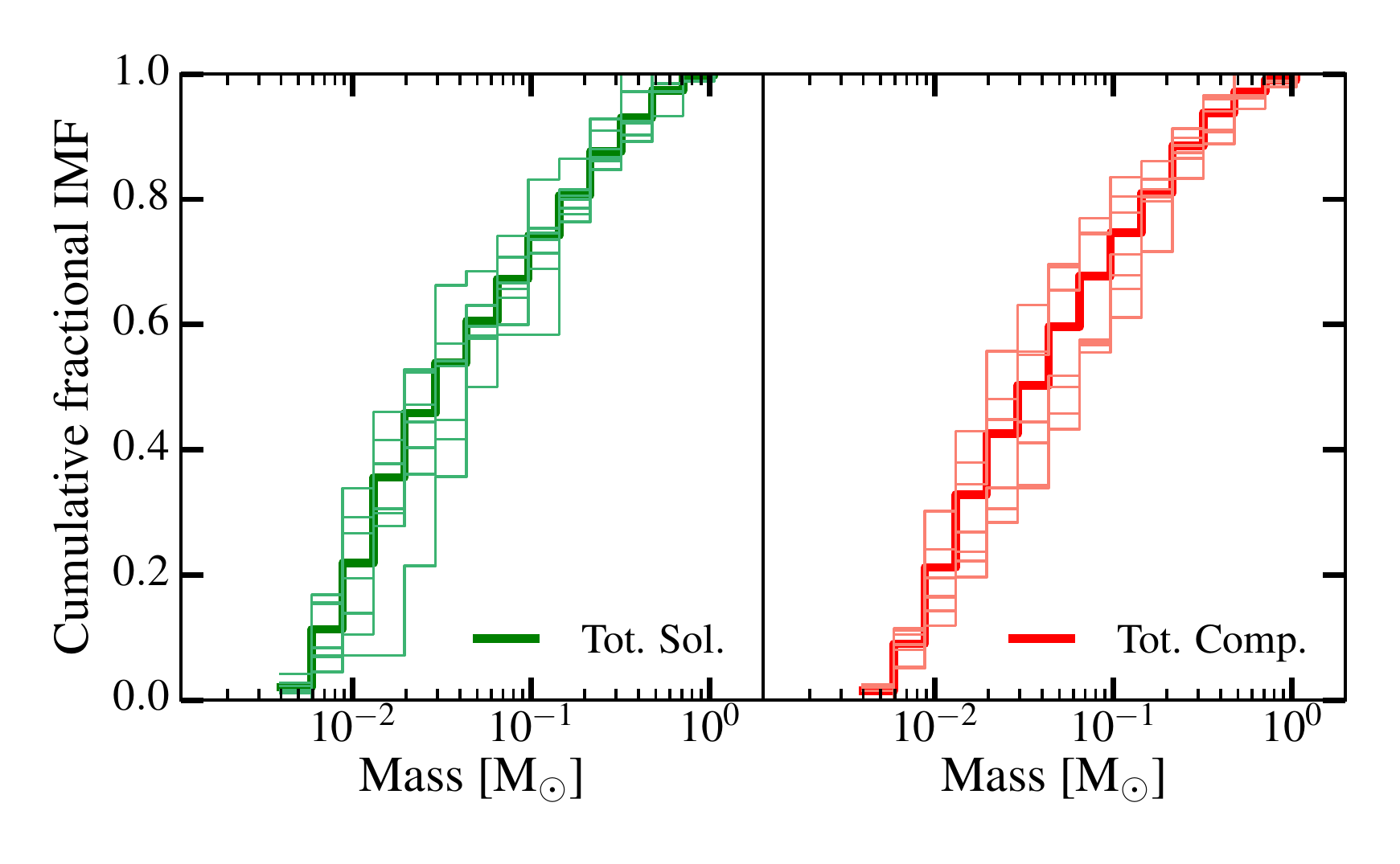}
\caption{Cumulative IMFs, comparing solenoidal (left) to compressive (right). Thick bold lines show the mean of all realisations while the thinner lines show the results from individual calculations.}
\label{fig:cumulativeIMFs}
\end{center}
\end{figure}

\subsection{Star formation rate}
 Figure~\ref{fig:sfr} shows the total stellar mass as a function of time, measured by the mass in sink particles. The onset of star formation occurs at $t \approx 0.2 t_{\rm ff}$ in the compressive case, compared to $t \approx 1.1 t_{\rm ff}$ in the solenoidal case. Once star formation starts in each calculation, the rate at which material is converted to stars is higher by a factor of $\sim 2$ in the compressive clouds compared to the solenoidal cloud.
 The overall efficiency of star formation is similar in both types of calculation over the time we have continued the simulations, with $\approx 15$\% of the gas mass converted to stars. However, the efficiency is higher on an absolute scale since this occurs over a shorter timescale in the compressive case. Also, the end of the simulations does not mark the end of the star formation process since the mass in stars continues to increase.


\subsection{Comparison of IMFs}
 Figure~\ref{fig:imfs} shows the IMFs from our simulations, combining all 7 realisations with solenoidal (left) and compressive driving (right), with the cumulative IMFs shown in Figure~\ref{fig:cumulativeIMFs}. The IMFs of stars that have finished accreting (235 of 388 and 298 of 533 sinks for solenoidal and compressive, respectively) are shown in red, while the IMF of all stars is shown in blue. The lowest mass possible in our calculations is $\approx 0.005 \text{ M}_\odot$ from the opacity limit for fragmentation, which sets the low-mass cutoff. The IMFs appear similar to those shown in \citetalias{bbb03} but with better statistics because of our multiple realisations. Our IMFs are also similar to those found by \citet{bate09a} from one calculation of a 500 $\text{M}_{\odot}$ cloud. In particular, we observe the statistically significant excess in low mass stars and brown dwarfs compared to the \citet{kroupa01} and \citet{chabrier05} IMFs (dashed and solid lines, respectively) that occurs when a barotropic equation of state is employed \citep[e.g.][]{bate09a,bate09}.

 There is no obvious difference between the IMFs produced by the different types of driving. Statistics confirm this --- a Kolmogorov-Smirnov test gives a p-value of 0.71 between the two distributions when considering all sink particles, and a p-value of 0.98 when considering only sinks that have finished accreting. This means we cannot reject the hypothesis that the samples come from the same underlying distribution. Thus, while the type of driving changes the density PDF, the resultant IMFs are indistinguishable.

\section{Discussion and Conclusions}
\label{sec:summary}
We presented the results of 14 numerical simulations of the gravitational collapse of 50~M$_{\odot}$ molecular clouds, each impulsively driven with a different random solenoidal or compressive velocity field to test the effect of the initial turbulence on the IMF. We resolved fragmentation to the opacity limit, at which point sink particles were inserted. By allowing the sink particles to accrete and grow in mass, we directly measured the masses of the resultant cluster of stars.

We found that while the initial turbulent velocity fields yielded different density PDFs during the initial collapse phase (before star formation begins), they had no significant effect on the IMF. However, the star formation rate was $\approx$ 2 times greater in the compressively driven clouds, with the onset of star formation occurring 0.9 free-fall times earlier. 
Our findings are consistent with \citet{girichidisetal11}, who found their IMFs unchanged by the ratio of solenoidal to compressive modes in the initial turbulence, and with \citet{bate09b} who found that using a different initial kinetic power spectrum did not significantly alter the resulting IMF.

The main caveat to our study is that we assumed impulsive turbulent driving, which does not produce a statistical steady state. Thus it may be argued that the turbulent support present in the collapsing cores has already decayed by the time star formation occurs.  Also, our density PDFs evolve in time and do not maintain the empirical relation between the variance, Mach number and the ratio of solenoidal and compressive modes (\eqnref{eq:sigma}; see Figure~\ref{fig:pdf_stats}). 
However, the decaying regime is important as it may better represent dense cores prior to star formation \citep[e.g.][]{ladaetal08} and thus driving of the velocity field by outflows and radiative feedback.

 The best answer to the above caveat is provided in the complementary study by \citet{bertelli-mottaetal16}. Although these authors did not resolve the IMF to the opacity limit, they used clouds driven to a statistical steady state inside a periodic box, before `switching on' gravity to collapse the cloud. Importantly, the turbulence in their experiments was continually driven throughout the calculations, producing PDFs which match \eqnref{eq:sigma}. Despite this, in their `high density' simulations which are most similar to ours, \citet{bertelli-mottaetal16} found no correlation between the properties of the turbulence and the resulting shape of the IMF, which is consistent with our findings. Furthermore, the trends found in their `low density' simulations, though too low resolution to probe the IMF directly, were also not consistent with the predictions of existing analytic theories. The authors attribute the null result in their `high density' simulations to the IMF being determined mainly by dynamical evolution of the fragments under the influence of self-gravity, which is also the case in our study. Thus, whether or not turbulence is driven or decaying, it would appear to have little or no influence on the IMF.

 Truly realistic simulations require an understanding of the physical source of turbulent driving in the interstellar medium. Our simulations also did not include radiative transfer or magnetic fields, both of which play an important role in determining the IMF. Furthermore, our ability to probe the IMF at $\text{M} \gtrsim 1 \text{M}_{\odot}$ is limited by the $50~\text{M}_{\odot}$ total mass of our model clouds. Worthwhile follow-up studies would include radiative feedback and more massive clouds \citep[e.g.][]{bate12,kkm12}, and magnetic fields \citep[e.g.][]{myersetal14}.
 

\section*{Acknowledgements}
We thank the anonymous referee for comments which have improved the paper. We acknowledge CPU time on gSTAR, funded by Swinburne University and the Australian Government. This project was funded via Australian Research Council Discovery Project DP130102078 and Future Fellowship FT130100034. We used \textsc{splash} \citep{price07}.

\label{lastpage}

\bibliography{dan}


\end{document}